\documentclass[twocolumn,showpacs,preprintnumbers]{revtex4}
\usepackage{amssymb}
\usepackage{graphicx}
\usepackage{dcolumn}
\usepackage{bm}

\begin{document}

\title{Reversable heat flow through the carbon nanotube junctions}
\author{S.~E.~Shafranjuk}
\homepage{http://kyiv.phys.northwestern.edu}
\affiliation{Department of Physics and Astronomy, Northwestern University, Evanston, IL
60208}
\date{\today }

\begin{abstract}
Microscopic mechanisms of externally controlled reversable heat flow through the carbon nanotube junctions (NJ) are studied theoretically. Our model suggests that the heat is transfered along the tube section ${\cal T}$ by electrons ($e$) and holes ($h$) moving ballistically in either in parallel or in opposite directions and accelerated by the bias source-drain voltage $V_{\rm SD}$ (Peltier effect). We compute the Seebeck coefficient $\alpha $, electric $\sigma$ and thermal $\kappa$ conductivities and find that their magnitudes strongly depend on $V_{\rm SD}$ and $V_{\rm G}$. The sign reversal of $\alpha$ versus the sign of $V_{\rm G}$ formerly observed experimentally is interpreted in this work in terms of so-called chiral tunneling phenomena (Klein paradox). 
\end{abstract}

\maketitle

Physics of the heat transfer determines functionality, precision and effectiveness of solid state nanocoolers\cite{NC1,NC2,NC3,Kim} which are environment friendly and have a lot of applications in the experimental physics, nanoelectronics, chemistry, industry and medicine. Therefore exploiting of new thermoelectric materials with high figures of merit $Z \cdot T$ ($T$ being the temperature) attracts a lot of attention. Recently such interest arose toward the carbon nanotube and graphene junctions which electronic properties are highly unconventional \cite{Kim}. The thermoelectric power experiments \cite{Kim} addressed single wall carbon nanotube junctions. In Ref. \cite{Kim} a temperature difference $\Delta T$ induced a finite bias voltage $\Delta V_p$ across the junction which sign changed versus  the gate voltage $V_{\rm G}$. A question here is how that unconventional thermoelectric behavior is related to the intrinsic nature of the carbon nanotube and graphene? It is widely accepted that the charge carrier motion in carbon nanotubes and in graphene is essentially phase-correllated. For such a reason the conducting electrons and holes in that materials behave as relativistic massless 'chiral fermions' (CF) characterized by a 'pseudospin' (see review \cite{Ando-rev} and references thereis). 
\begin{figure}[tbp]
\includegraphics{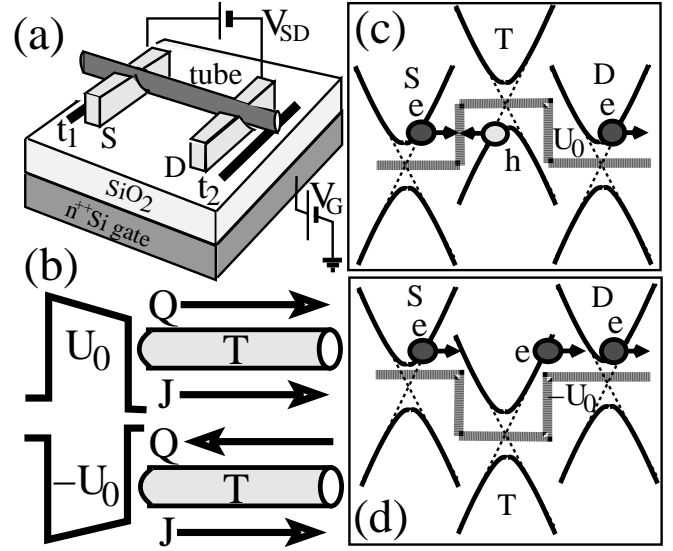}
\caption{(a) Typical carbon nanotube junction (NJ) controlled by the gate voltage $V_{\mathrm{G}}$. The source (S) and drain (D) electrodes are attached to the nanotube section ${\cal T}$, $t_{1,2}$ are metallic stripes serving as termometers. (b)~Schematics of the heat (Q) and electric (J) transport responsible for the reversable Peltier effect. (c,d)~Energy diagrams of basic microscopic tunneling processes involving electrons ($e$) and holes ($h$) across the chiral barrier (c) and well (d).}
\label{fig:Setup_a}
\end{figure}

In this Letter we argue that the phase-correlated thermoelectric transport of charge carriers implicates a voltage-controlled and reversable  heat flow through the single wall carbon nanotube junctions (Peltier effect). The enhancement of $Z \cdot T$ in those 1D devices where  the charge carriers propagate ballistically occurs due to strong van Hove singularities (VHS). The VHS position is tuned by the gate voltage $V_{\rm G}$ in respect to the Fermi level $\varepsilon_{\rm F}$ of the electrodes. When an VHS and $\varepsilon_{\rm F}$ match each other, it results in a sufficient density of charge carriers which contribute to the electric conductivity despite the Fermi energy itself is relatively small. A finite gate voltage $V_{\rm G} \neq 0$ is not just merely supplies either the electrons or holes into ${\cal T}$, but rather creates a potential barrier ($U_0>0$)  or well ($U_0<0$) for chiral fermions transmitted across the junction as illustarted in Figs.~\ref{fig:Setup_a}(b-d). A finite sorce-drain d.c. bias voltage $V_{\mathrm{SD}}\neq 0$ applied across the junction tilts the chiral barrier and causes the electric charge carriers to accelerate. This means that the energy of the charge carriers changes by $\delta =eV_{\mathrm{SD}}$, which inflicts a local temperature change $\Delta T$ at the corresponding junction$^{\prime }$s end. The sign of $\delta $ (and hence of $\Delta T$) depends on whether the charge carriers inside the barrier are electrons or holes. Other thermal characteristics of the carbon nanotube and graphene junctions depend on precise microscopic mechanisms acting on the nanoscale. The setup shown in Fig.~ \ref{fig:Setup_a} allows controlling the sign of $\Delta T$ by merely flipping the sign of the gate voltage [see the sketch in Fig. ~\ref{fig:Setup_a}(b)]. The source (S) and drain (D) electrodes can either be metallic or made of the nanotube/graphene itself. The potential chiral barrier $U_0$ is induced by the gate voltage $V_{\mathrm{G}}$ from a Si gate.  If $V_{\mathrm{G}}>0$, the transport mechanism is the chiral tunneling (CT). It presumes a constructive quantum interference between an incoming electron ($e$) and a hole ($h$) moving inside the graphene barrier in a reverse direction and is characterized by the same pseudospin as sketched in Fig.~\ref{fig:Setup_a}(c) (Klein paradox\cite{Strange,Krekora}). The interference pattern resulting from the chiral tunneling is very sensitive to the phase difference $\varphi $ between the $e$ and $h$ wavefunctions.  During the CT process the electric current is transferred by holes, which ballistically propagate in a reverse direction. That means the heat and the electric current are directed in antiparallel. When the gate voltage is reversed ($V_{\mathrm{G}} < 0$), the heat and the electric current flow in parallel, since both of them are transferred by electrons. This reverses the heat flow along ${\cal T}$ as compared to the former case $V_{\rm G} > 0$. We emphasize that the reversable heat flow originates from the phase correlation between the electrons and holes which constitutes its quantum mechanical origin. Our finding suggest that the thermoelectric transport serves as another independent probe of the phase-coherence in the carbon nanotube junctions.

\begin{figure}[tbp]
\includegraphics{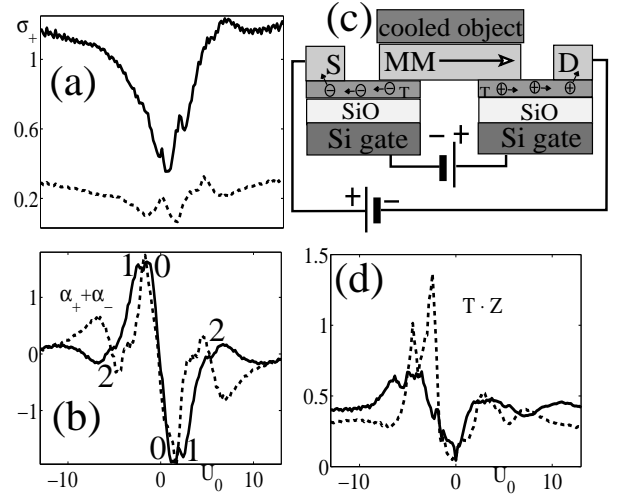}
\caption{(a)~The d.c. electric conductivity $\sigma_{+} (U_0)$ [in units of $e^{2}v_{F}N\left(
0\right) $] for the two SWCNT diameters $d_{\rm T}=0.75$ (dashed curve) and $d_{\rm T}=1.5$  (solid curve) computed at $T=0.5$ and $\nu=1$.  (b)~The net Seebeck coefficient $\alpha_{+}+\alpha_{-}$ [in units of $1/\left( k_{\rm  B}TeV_{\mathrm{SD}}\right)$] versus $U_0$ for the same SWCNTs at $T=0.5$ and $\nu=1$. Indices 0, 1, 2 in (b) mark features coming from the electron energy subbands with $n=0,1,2$ correspondingly. (c)~Schematics of the reversable Peltier nanocooler combined of two NJs [see the former Fig. 1(a)] with individual gate voltage $V_{\rm G}^{(1,2)}$ control each. (d) ~The corresponding dependence for the figure of merit $T \cdot Z (U_0)$. }
\label{fig:Fig_mer_U0}
\end{figure}

Thermoelectric transport properties of the Peltier coolers are characterized by the figure of merit $Z \cdot T$. For a symmetric setup shown in Fig. ~\ref{fig:Fig_mer_U0}(c) $Z$ takes the form
\begin{equation}
Z=\frac{\left( \alpha _{+}-\alpha _{-}\right) ^{2}\sigma _{\rm{seq}}}{%
\kappa _{+}+\kappa _{-}+\kappa _{L}}  \label{Z_M}
\end{equation}
In Eq. (\ref{Z_M}) the lattice thermal conductivity is $\kappa _{L}=C_{L}v_{L}l_{L}/3$, $C_{L}$ is the graphene heat capacity, which at $T=300 K$ is $C_{L} = 8.5$ J/(mol$\cdot $K), $v_{L} \approx 0.1 v$ is the graphene sound velocity, where  $v=8.1\cdot 10^{5}$ m/s is the Fermi velocity in graphene, $l_{L}$ is the phonon free path inside the nanotube. A most conservative estimation used here implies that $l_{L}$ is comparable with the tube diameter, i.e., $l_{L} \approx d_{\rm T}$. The Seebeck coefficient of electron (+) and holes (-),  $\alpha _{\pm}$, is obtained as $\alpha _{\pm}=-1/(q_{\pm} T) (L^{\left( 1\right) }_{\pm}/L^{\left(2\right) }_{\pm})$ , where $q_{\pm}={\mp }e$ is the electric charge of the electron (hole). Besides in Eq.~(\ref{Z_M}) $\sigma _{\rm{seq}}$ is the conductivity of two equal size conducting tubes ${\cal T}$ connected in a sequence, $\sigma _{\rm{seq}}=\sigma _{+}\sigma _{-}/\left( \sigma _{+}+\sigma _{-}\right) $ where $\sigma _{\pm}=L_{\pm}^{\left( 0\right) }$ and $\kappa _{n,p}$ is the electron (hole) thermal conductivity $\kappa _{\pm} =\left( L_{\pm}^{\left( 2\right) }-\left[
L_{\pm}^{\left( 1\right) }\right] ^{2}/L_{\pm}^{\left( 0\right) }\right) /(e^{2}T)$.
The functions $L_{\pm}^{\left( \alpha \right)}$ entering the above formulas are
\begin{equation}
L_{\pm}^{\left( \alpha \right) }=2e^{2}N\left( 0\right) \int \left(
\varepsilon -V_{\rm SD} \right) ^{\alpha }\left\vert t_{\varepsilon
}^{\pm}\right\vert ^{2}\left( -\frac{\partial n_{\varepsilon }}{\partial
\varepsilon }\right) d\varepsilon  \label{L_p}
\end{equation}%
The functions $%
L_{\pm}^{\left( \alpha \right) }$ are computed from a microscopic model. If one neglects by detailed energy dependence of $t_{\varepsilon }^{\pm}$ and merely sets $\left\vert t_{\varepsilon }^{\pm}\right\vert ^{2}\rightarrow \bar{T}$, 
then $\sigma _{p,n}=e^{2}N\left( 0\right) \bar{T}$ where $N\left( 0\right) $ is the electron density
of states at the Fermi level of electrodes, and
\begin{equation}
\alpha _{\pm} =-\frac{6}{q_{\pm} T}\frac{T\log 2-V_{\rm SD} /2}{\pi ^{2}T^{2}+3V_{\rm SD} (V_{\rm SD} -4T\log  2)}
\label{alpha0}
\end{equation}
and the electron (hole) thermal conductivity $\kappa _{n,p}$ becomes%
\begin{eqnarray}
\kappa _{\pm}=N\left( 0\right) \bar{T}(
\pi ^{2}T^{2}/3+V_{\rm SD} (V_{\rm SD} -4T\log 2)  \nonumber \\
-4\left( T\log 2-V_{\rm SD} /2 \right) ^{2})/T
\label{kappa0}
\end{eqnarray}
The estimations using Eqs. (\ref{Z_M})-(\ref{kappa0}) give the figure of merit for a symmetric setup as%
\begin{eqnarray}
T \cdot Z &=&\frac{288}{ \alpha_Z ^{2}}\frac{ ( \log 2-V_{\rm SD} /2) ^{2}}{\beta_Z+e^{2}T\cdot \text{$\kappa _{L}$}
} \label{ZT}
\end{eqnarray}%
where $\alpha_Z=\pi ^{2}T^{2}+3V_{\rm SD} (V_{\rm SD} -4T\log 2)$, $\beta_Z = ( \pi ^{2}T^{2}+3V_{\rm SD} (V_{\rm SD} -4T\log 2)) /3-4( T\log 2-V_{\rm SD} /2) ^{2} $. One may notice that $T\cdot Z$ does not depend on the properties of electrodes but instead strongly depends on $T$ and $V_{\rm SD}$. For $\kappa _{L}=10$ the product  $T \cdot Z$ may be increased by 2 orders of magnitude when neglecting by the phonon heat conductivity. In Eq. (\ref{alpha0})-(\ref{ZT})  we used that $L_{\pm}^{\left( 0\right) }=e^{2}N\left( 0\right) \bar{T} $ $L_{\pm}^{\left( 1\right) }=2e^{2}N\left( 0\right) \bar{T}\cdot \left( T\log 2-V_{\rm SD} /2 \right) $, and $ L_{\pm}^{\left( 2\right) }=e^{2}N\left( 0\right) \bar{T}\left( \pi ^{2}T^{2}/3+V_{\rm SD} (V_{\rm SD} -4T\log 2)\right) $. The above rough estimations illustrate the temperature behaviour of the figure of merit disregarding however the microscopic mechanisms of the thermoelectric transport. A deeper understanding is achieved with using of a microscopic model where the electron and hole envelope wave functions satisfy the Dirac equation.  That model allows computing the transparency  $\bar{T}=\left\vert t_{\varepsilon }^{\pm}\right\vert ^{2}$ of the chiral barrier and well shown in Fig. \ref{fig:Setup_a}(b-d). We calculate the steady state thermoelectric characteristics of an NJ with metallic electrodes controlled by the gate voltage $V_{\mathrm{G}}$ and biased by the source-drain voltage $V_{\mathrm{SD}}$. In particular we will see that the gate voltage reverses the heat flow along ${\cal T}$, which is well consistent with the earlier experiments\cite{Kim}. The figure of merit $Z \cdot T$ strongly increases at certain magnitudes of $V_{\rm G}$ and $V_{\rm SD}$. The source drain bias voltage $V_{\mathrm{SD}}$ in our setup drops entirely on the nanotube [see its profile below Fig. 1(b)], which corresponds to a trapezoidal shape of the chiral barrier tilted proportionally to $V_{\mathrm{SD}}$. For the NJ having finite dimensions, the motion of CF is quantized. The quantization imposes additional constrains on the directional tunneling diagram. The Dirac equation for fermions is written as%
\begin{equation}
-i\hbar v \left( \left(\hat \sigma _{x}\otimes \hat 1\right) \partial
_{x}+\left(\hat \sigma _{y}\otimes \hat \tau _{z}\right) \partial
_{y}\right) \Psi +eU\left( x\right) \Psi =\varepsilon \Psi  \label{Dirac_A}
\end{equation}%
where $v=c/300$ is the massless fermion speed, $\varepsilon $ is the CF energy, $\hat \sigma _{i}$ and $\hat \tau_k$ are the Pauli matrices, $\{i,k\}=1\dots3$, the barrier potential $U(x)$ is induced by the
gate voltage $V_{\mathrm{G}}$. In the steady state, when the EF is off, the
electric current is fully suppressed when $V_{\mathrm{SD}}<U_{0}$ (for
typical gate voltage $V_{\mathrm{G}}=1$ V and the SiO$_2$ thickness $d=300$
nm one finds\cite{Japan2} $U_0=2$ meV). The CF energy in the
leads ($x<0$ and $x>L$) reads $\varepsilon =eV_{\infty }\pm \hbar v\sqrt{%
k^{2}+q_{\nu}^{2}(n)}=\hbar v\left( k_{Fm}\pm \sqrt{k^{2}+q_{\nu}^{2}(n)}\right) $ where $k_{Fm}$ is the Fermi wavevector in the S and D electrodes. Then in the electrodes one takes $k=\sqrt{\left( \varepsilon/\hbar
v -k_{Fm}\right) ^{2}-q_{\nu }\left( n\right) ^{2}}$.
Inside the tube one uses $\tilde{k}=\sqrt{\left(( \varepsilon -eU_{0})/\hbar
v\right) ^{2}-q_{\nu }\left( n\right) ^{2}}$, where the electron wave vector
in the transverasal direction is  $q_{\nu }\left( n\right) =\left( 2
/d_{\rm T}\right) \left( n-\nu /3\right) $, $d_{\rm T}$ is the nanotube diameter, $n$ is the electron subband index, $\nu=n+m-3N$, $n$, $m$, and $N$ are integer numbers  related to the translation vector ${\cal L}=n {\bf a}+m{\bf b}$ of the graphene lattice, ${\bf a}$ and ${\bf b}$ are primitive translation vectors. In particular index  $\nu =0$ when the $[m,n]$ tube is metallic while $\nu \neq 0$ for semiconducting and dielectric tubes. 

In a simplest case $V_{\rm SD}=0$ from continuous boundary conditions one gets
\begin{equation}
t_{n}=2e^{-i\varphi }k \tilde{k}s\tilde{s}/\tilde{\cal D}
\label{tr_0}
\end{equation}
where $\tilde{\cal D}=2k \tilde{k} s\tilde{s}\cos (\tilde{k}L)-i\Theta \sin (\tilde{k}L)$, $L$ is the tube length, $\tilde{\kappa }= \sqrt{\tilde{k}^{2}+q^{2}}$, $\Theta = \kappa\tilde{\kappa }(s^{2}+\tilde{s}^{2})-2s\tilde{s}q^{2}$, $\kappa= \sqrt{k^{2}+q^{2}}$, $s=1$, $\tilde{s}=\mathrm{sign}\left( \varepsilon -eU_{0}\right) $.
The transmission amplitude $t_{\varepsilon }$
is computed using a simple approach, which represents the CF envelope
wavefunctions $\hat \Psi $ as plane waves for a rectangular barrier and via
Airy functions\cite{Korn} for a trapezoidal barrier. In particular, the scattering
state for a trapezoidal barrier is constructed as 
\begin{eqnarray}
\Psi =\theta ( -x) [ \chi _{n,k}\mathrm{Ai}(k,x)+r_{n}\chi _{n,-k}\mathrm{Bi}%
(k,x)]  \nonumber \\
+\theta ( x-L) t_{n}\chi _{n,k}\mathrm{Ai}(k, x-L)+\theta ( x) \theta ( L-x)
\nonumber \\
\times [ \alpha _{n}\chi _{n,\tilde{k}}\mathrm{Ai}(\tilde{k}, x)e^{i\varphi
} +\beta _{n}\chi _{n,-\tilde{k}}\mathrm{Bi}(\tilde{k},x)e^{i\varphi }]
\end{eqnarray}
where $\mathrm{Ai}(k,x)$ and $\mathrm{Bi}(k,x)$ are the Airy functions, and
we also introduced the auxiliary functions $\chi _{n,k}\left( y\right)
=a_{n}\left\vert \uparrow \right\rangle \otimes \left( \left\vert \uparrow
\right\rangle +z_{n,k}\left\vert \downarrow \right\rangle \right)
e^{iq_{n}y} $ $+a_{n}^{\prime }\left\vert \downarrow \right\rangle \otimes
\left( z_{n,k}\left\vert \uparrow \right\rangle +\left\vert \downarrow
\right\rangle \right) e^{iq_{n}y}$ $+b_{n}\left\vert \uparrow \right\rangle
\otimes \left( z_{n,k}\left\vert \uparrow \right\rangle +\left\vert
\downarrow \right\rangle \right) e^{-iq_{n}y}$ $+b_{n}^{\prime }\left\vert
\downarrow \right\rangle \otimes \left( \left\vert \uparrow \right\rangle
+z_{n,k}\left\vert \downarrow \right\rangle \right) e^{-iq_{n}y}$ where $%
\otimes $ means the Kronecker product, $\left\vert \uparrow \right\rangle ^{T}=\left( 
\begin{array}{cc}
1 & 0%
\end{array}%
\right) $ and $\left\vert \downarrow \right\rangle ^{T}=\left( 
\begin{array}{cc}
0 & 1%
\end{array}%
\right) $ are 1$\times $2 matrices, $z_{n,k}=\pm \left( k+iq_{n}\right) /%
\sqrt{k^{2}+q_{n}^{2}}$, where $\pm $ signs apply to conductive (valence)
bands, $k$ is positive for conductance band and negative for the valence
band, the factor $z_{n,k}$ satisfies the identity $z_{n,k}z_{n,-k}=-1$.
Then one implements the above Eqs. (\ref{Z_M})-(\ref{L_p}), and (\ref{tr_0})  to obtain $\sigma _{\pm}$, $\alpha _{\pm}$, $\kappa_{\pm}$, and $T \cdot Z$.

\begin{figure}[tbp]
\includegraphics{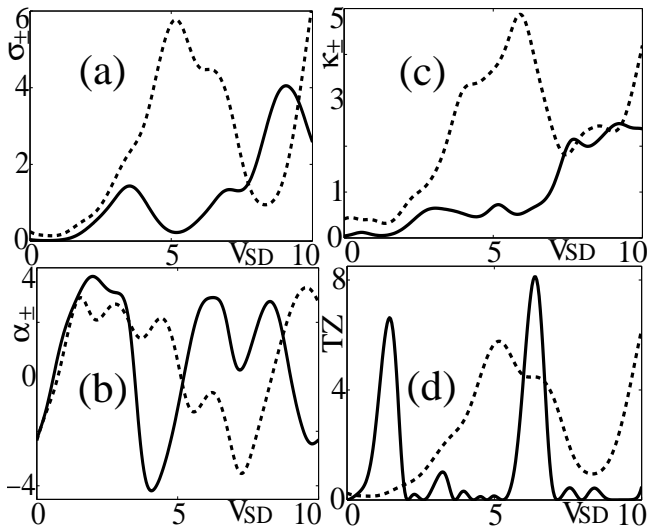}
\caption{The electric conductivity $\sigma_{\pm}$, Seebeck coefficient $\alpha_{\pm}$, the thermal conductivity $\kappa_{\pm}$, and the figure of merit $Z $ of the reversable Peltier nanocooler at $T=0.5$ versus the source-drain voltage. All the units are the same as in former Fig. 2.}
\label{fig:Fig_mer_VSD}
\end{figure}

Our theoretical model interpretes the experimental data of Ref.~\cite{Kim} in terms of chiral tunneling. We find that the electric and heat transport characteristics of the nanotube junctions are pretty much determined by details of their electron spectrum. In particular, the electric and heat transport characteristics of the semiconducting nanotube (with $\nu=1$) junctions shown in Fig. ~\ref{fig:Fig_mer_U0} for two SWCNT diameters $d_{\rm T}=0.75$ (dashed curves) and $d_{\rm T}=1.5$  (solid curves) computed at $T=0.5$ remarkably depend on the barrier height $U_0$. The d.c. electric conductivity $\sigma_{+} (U_0)$ plotted in units of $e^{2}v_{F}N\left(
0\right) $ in Fig.~\ref{fig:Fig_mer_U0}(a) shows a huge dip in vicinity of  $U_0 \approx 0$, which comes from the dispersion law $\tilde{k}\propto \sqrt{\left( eU_{0}/\hbar v\right)^{2}-(2 \nu/3 d_{\rm T})^{2}}$ in the lowest subband with $n=0$. The upper electron subbands with $n=1,2,3$ give a fine structure in the transport characteristics which is remakably pronounced in all the $\sigma_{\pm} (U_0)$, $\alpha_{\pm}(U_0)$, $\kappa_{\pm}(U_0)$, and $T \cdot Z (U_0)$ curves, as is evident from Fig.~\ref{fig:Fig_mer_U0}. To distinguish the contributions from different electron subbands with $n=0,1,2$ we mark the corresponding features by indices 0,1,2 in Fig.~\ref{fig:Fig_mer_U0}(b). The positions and magnitudes of those features reflect the electron band structure of the tube or the graphene stripe. Similar fine structure features coming from different electron subbands are visible also in Figs.~\ref{fig:Fig_mer_U0}(a), (c,d) for two the SWCNT diameters $d_{\rm T}=0.75$ (dashed curve) and $d_{\rm T}=1.5$  (solid curve) computed at $T=0.5$ and $\nu=1$.  The reversable heat flow due to the chiral tunneling is well pronounced in the Seebeck coefficient $\alpha_{\pm}$. In Fig.~\ref{fig:Fig_mer_U0}(b) we plot the net  Seebeck coefficient $\alpha_{+}+\alpha_{-}$ [in units of $1/\left( k_{\rm B}T e V_{\mathrm{SD}}\right)$]. One see that $\alpha_{+}+\alpha_{-}$ reverses its sign as the sign of $U_0$ changes. This sign change along with the specific one-dimensional bandstructure of the single wall carbon nanotubes can be exploited for creating of very efficient Peltier nanocoolers. An example of that cooler setup is sketched in Fig.~\ref{fig:Fig_mer_U0}(c). The reversable Peltier nanocooler is combined of two NJs [see the former Fig. 1(a)] with individual gate voltage $V_{\rm G}^{(1,2)}$ control for each of the tube. A significant enhancement of the figure of merit $Z \cdot T$ in those 1D devices is achieved by matching of the van Hove singularities (VHS) with position of the Fermi level $\varepsilon_{\rm F}$ in the electrodes by changing the gate voltage $V_{\rm G}$ and the bias voltage $V_{\rm SD}$. The aforementioned peculiarities coming from the chiral tunneling and from the electron band structure actually determine behavior of $T \cdot Z$ versus $U_0$ shown in Fig.~\ref{fig:Fig_mer_U0}(d). From that Fig.~\ref{fig:Fig_mer_U0}(d) one can see that for the semiconducting ($\nu=1$) tube with diameter $d=0.75$ the product  $T \cdot Z$ may exceed 1. Further improvement of the $T \cdot Z$ product about by one order of magnitude is acomplished by applying a finite bias voltage $V_{\rm SD}$ as shown in plots of $\sigma_{\pm} (V_{\rm SD})$, $\alpha_{\pm}(V_{\rm SD})$, $\kappa_{\pm}(V_{\rm SD})$, and $T \cdot Z (V_{\rm SD})$ in  Fig.~\ref{fig:Fig_mer_VSD}(a-d). The transport coefficients were computed at the temperature $T=0.3$ (in units $U_0$, which for $U_0=50$ mV corresponds $T=150$ K). One may notice that the $T \cdot Z$ product also depends on the tube diameter $d_{\rm T}$. For a thinner tube with $d=0.75$ one gets $T \cdot Z=7$ already at $V_{\rm SD}=1.2$ while for the thicker tube $d=1.5$  one achieves $T \cdot Z=8$ by applying a much higher bias voltage $V_{\rm SD}=6.5$. Similar double peak structure $T\cdot Z(V_{\rm SD})$ coming from matching of the VHSs with $\varepsilon_{\rm F}$ of electrodes [see solid curve in Fig.~\ref{fig:Fig_mer_VSD}(d)] had also been proclaimed in Bi nanowires \cite{Dress}. More significant increase of the $T \cdot Z$ product is obtained at lower temperatutes. Using the cooler setup shown in Fig.~\ref{fig:Fig_mer_U0}(c) one may potentially get even $T \cdot Z=10^2-10^3$ for  temperatures $T=10$K down to 1 K and with an appropriate tuning by $V_{\rm G}$ and $V_{\rm SD}$ simultaneously.
 Inclusion of the Shottky barriers and Coulomb blockade in the present model is pretty straiforward and will be given elsewhere.
 Our study indicates that narrow semiconducting single wall CNTs are more efficient in coolers than metallic CNT and graphene stripes (where there is no VHS for the lower subband $n=0$). An evident reason is that the matching of VHSs with $\varepsilon_{\rm F}$ in electrodes results in a better cooling performance. Another strong benefit of the NJ cooler is that one avoids a mutual compensation of  $\alpha_{+}$ and $\alpha_{-}$ in Eq. \ref{Z_M} as it happens, e.g., for Bi nanowire cooler (see Ref.~\cite{Dress}) since one appropriately flips signs of $\alpha_{\pm}$ merely by changing $V_{\rm G}$.
 
In conclusion we suggested a theoretical model of the electric and heat transport across the carbon nanotube junction. We find that the heat transport strongly depends on the polarity and magnitude of the source-drain and gate voltages. 
That voltages tune the basic transport properties of the carbon nanotubes via affecting of their electronic structure and phase-correlated transport of charge carriers. Due to resonant character of chiral tunneling and low inelastic scattering rates the heat current density is typically much higher than in ordinary semiconducting devices. At the same time, the Peltier effect is well pronounced up to relatively high temperatures. The reversable Peltier effect shows great promises for cooling nanodevices and termometers.

I wish to thank H.~Weinstock, V.~Chandrasekhar, P.~Barbara, A.~Sergeev and H.~Espinosa for fruitful discussions.

\end{document}